\documentclass[11pt,a4paper]{article}

\usepackage[english]{babel}
\usepackage{ifthen}
\usepackage{xspace}
\usepackage[utf8]{inputenc}
\usepackage{microtype}
\usepackage{xspace}
\usepackage{enumitem}
\usepackage{geometry}
\usepackage{amsmath,amssymb}
\usepackage{nicefrac}
\usepackage{array,dsfont}
\usepackage{pgf}
\usepackage{pgfkeys}
\usepackage{pgfplots}
\usepackage{graphicx,tabularx}
	\graphicspath{{Images/}}
\usepackage{subfig}
\usepackage{here}
\usepackage{pifont}
\usepackage{rotating}
\usepackage{booktabs}
\usepackage{url}
\usepackage[colorlinks, pdfpagelabels, pdfstartview = FitH, bookmarksopen = true, bookmarksnumbered = true, linkcolor = black, plainpages = false, hypertexnames = false, citecolor = black] {hyperref}

\addto{\captionsenglish}{

}

\newcommand{\orcidPS}{\href{https://orcid.org/0000-0002-5976-0317}{\includegraphics[height=9pt]{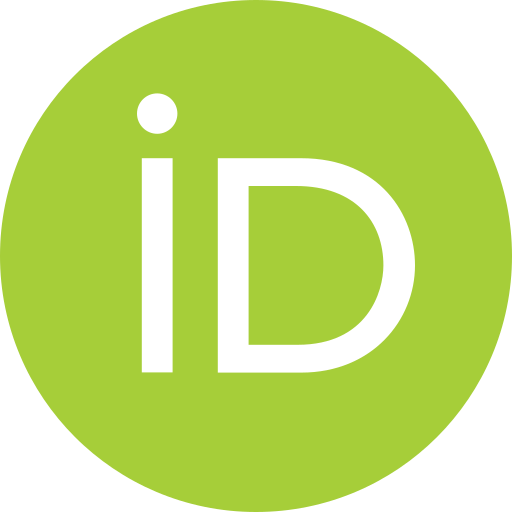}~}}
\newcommand{\orcidJR}{\href{https://orcid.org/0000-0003-1866-0157}{\includegraphics[height=9pt]{Logos/orcid.png}~}}

\newcommand{\whizard}{\texttt{WHIZARD}\xspace}
\newcommand{\whizardtwo}{\texttt{WHIZARD2}\xspace}
\newcommand{\OMega}{\texttt{O'Mega}\xspace}
\newcommand{\vamp}{\texttt{VAMP}\xspace}
\newcommand{\vamptwo}{\texttt{VAMP2}\xspace}
\newcommand{\circe}{\texttt{CIRCE}\xspace}
\newcommand{\circeone}{\texttt{CIRCE1}\xspace}
\newcommand{\circetwo}{\texttt{CIRCE2}\xspace}
\newcommand{\guineapig}{\texttt{GuineaPig}\xspace}

\newcommand{\openloopstwo}{\texttt{OpenLoops2}\xspace}

\newcommand{\gosam}{\texttt{Gosam}\xspace}

\newcommand{\powheg}{\texttt{POWHEG}\xspace}
\newcommand{\recola}{\texttt{Recola}\xspace}
\newcommand{\madgraph}{\texttt{MG5\_aMC}\xspace}

\newcommand{\sherpa}{\texttt{Sherpa}\xspace}

\newcommand{\GeV}{\ensuremath{\,\text{GeV}}\xspace}
\newcommand{\TeV}{\ensuremath{\,\text{TeV}}\xspace}

\newcommand{\bbar}{\overline{b}}
\newcommand{\tbar}{\overline{t}}
\newcommand{\nubar}{\overline{\nu}}

\newcommand{\eett}{$e^+ e^- \to t \tbar$\xspace}
\newcommand{\eetth}{$e^+ e^- \to t \tbar h$\xspace}
\newcommand{\eejjjjj}{$e^+ e^- \to jjjjj$\xspace}
\newcommand{\eejjjjjj}{$e^+ e^- \to jjjjjj$\xspace}

\makeatletter
\newif\if@preliminary
\@preliminaryfalse
\def\preliminary{\@preliminarytrue}

\def\preprintno#1{\def\@preprintno{#1}}
\def\address#1{\def\@address{#1}}
\def\email#1#2{\thanks{\tt #1@{}#2}}
\def\abstract#1{\def\@abstract{#1}}
\renewcommand\abstractname{ABSTRACT}
\newlength\preprintnoskip
\newlength\abstractwidth
\@titlepagetrue
\renewcommand\maketitle{\begin{titlepage}%
  \let\footnotesize\small
  \hfill\parbox{\preprintnoskip}{%
  \begin{flushright}\@preprintno\end{flushright}}\hspace*{1cm}
  \vskip 60\p@
  \begin{center}%
    {\Large\bf\boldmath \@title \par}\vskip 1cm%
    {\sc\@author \par}\vskip 3mm%
    {\@address \par}%
    \if@preliminary
      \vskip 2cm {\large\sf PRELIMINARY DRAFT \par \@date}%
    \fi
  \end{center}\par
  \@thanks
  \vfill
  \begin{center}%
    \parbox{\abstractwidth}{\centerline{\abstractname}%
    \vskip 3mm%
    \@abstract}
  \end{center}
  \end{titlepage}%
  \setcounter{footnote}{0}%
  \let\thanks\relax\let\maketitle\relax
  \gdef\@thanks{}\gdef\@author{}\gdef\@address{}%
  \gdef\@title{}\gdef\@abstract{}\gdef\@preprintno{}
}%
\def\@citex[#1]#2{\if@filesw\immediate\write\@auxout{\string\citation{#2}}\fi
  \def\@citea{}\@cite{\@for\@citeb:=#2\do
    {\@citea\def\@citea{,\penalty\@m}\@ifundefined
       {b@\@citeb}{{\bf ?}\@warning
       {Citation `\@citeb' on page \thepage \space undefined}}%
\hbox{\csname b@\@citeb\endcsname}}}{#1}}
\def\citerange{\@ifnextchar [{\@tempswatrue\@citexr}{\@tempswafalse\@citexr[]}}
\def\@citexr[#1]#2{\if@filesw\immediate\write\@auxout{\string\citation{#2}}\fi
  \def\@citea{}\@cite{\@for\@citeb:=#2\do
    {\@citea\def\@citea{--\penalty\@m}\@ifundefined
       {b@\@citeb}{{\bf ?}\@warning
       {Citation `\@citeb' on page \thepage \space undefined}}%
\hbox{\csname b@\@citeb\endcsname}}}{#1}}
\long\def\@makecaption#1#2{%
  \sbox\@tempboxa{#1: \emph{#2}}%
  \ifdim \wd\@tempboxa >\hsize
    #1: \emph{#2}\par
  \else
    \hbox to\hsize{\hfil\box\@tempboxa\hfil}%
  \fi
  \vskip\belowcaptionskip}
\def\fmslash{\@ifnextchar[{\fmsl@sh}{\fmsl@sh[0mu]}}
\def\fmsl@sh[#1]#2{%
  \mathchoice
    {\@fmsl@sh\displaystyle{#1}{#2}}%
    {\@fmsl@sh\textstyle{#1}{#2}}%
    {\@fmsl@sh\scriptstyle{#1}{#2}}%
    {\@fmsl@sh\scriptscriptstyle{#1}{#2}}}
\def\@fmsl@sh#1#2#3{\m@th\ooalign{$\hfil#1\mkern#2/\hfil$\crcr$#1#3$}}
\makeatother

\expandafter\def\csname pgf@textdist@protect\endcsname{}
\input pgfmath.tex
\pgfset{
    number format/.cd,
    min exponent for 1000 sep=4,
    int detect,
 }

\newcommand{\kfactor}[2]{%
 \def\x{#1}%
 \def\y{#2}%
 \def\xy{(\x) / (\y)}%
 \pgfmathparse{\xy}%
 \pgfmathprintnumber[fixed,zerofill,precision=3]{\pgfmathresult}%
 \ignorespaces%
}

\newcommand{\tableline}[5]{
 \ifthenelse{\equal{#1}{}}
   {\unskip\na & \na & \na}
   {\ifthenelse{\equal{#3}{}}
     {
       \ifthenelse{\equal{#5}{}}
       {\unskip$#1(#2)$ & \na & \na}
       {\unskip$#1(#2) \cdot 10^{#5}$ & \na & \na}
           }
     {
       \ifthenelse{\equal{#5}{}}
       {\unskip$#1(#2)$ & $#3(#4)$ & \kfactor{#3}{#1}}
       {\unskip$#1(#2) \cdot 10^{#5}$ & $#3(#4) \cdot 10^{#5}$ & \kfactor{#3}{#1}}
     }
         }
}

\newcommand{\stddev}[4]{%
 \def\ina{#1}%
 \def\inb{#3}%
 \def\erra{#2}%
 \def\errb{#4}%
 \pgfkeys{/pgf/fpu=true}
 \def\absdiff{abs((\ina) - (\inb))}%
 \def\errfunc{sqrt((\erra) * (\erra) + (\errb) * (\errb))}%
 \def\stddevval{(\absdiff) / (\errfunc)}%
 \pgfmathparse{\stddevval}%
 \pgfmathprintnumber[fixed,zerofill,precision=2]{\pgfmathresult}%
 \pgfkeys{/pgf/fpu=false}
 \ignorespaces%
}

\usepackage[
style=numeric,
citestyle=numeric-comp,
url=false,
giveninits=true,
sorting=none,
backend=bibtex
]{biblatex}
  \renewbibmacro{in:}{}
  \bibliography{literature.bib}

\begin{document}
\date{\today}
\preprintno{DESY 21-050}
\title{\whizard \texttt{3.0}: Status and News}
\author{Pascal Stienemeier\orcidPS%
  \footnote{Talk presented at the International Workshop on Future Linear Colliders (LCWS 2021), 15-18 March 2021, PD1-4.}%
  $\!^,$%
  \email{pascal.stienemeier}{desy.de}$^a$;
  Simon Braß\email{simon.brass}{desy.de}$^a$,
  Pia Bredt\email{pia.bredt}{desy.de}$^a$,
  Wolfgang Kilian\email{kilian}{physik.uni-siegen.de}$^b$,
  Nils Kreher\email{kreher}{physik.uni-siegen.de}$^b$,
  Thorsten Ohl\email{ohl}{physik.uni-wuerzburg.de}$^c$,
  Jürgen Reuter\orcidJR\email{juergen.reuter}{desy.de}$^a$,
  Vincent Rothe\email{vincent.rothe}{desy.de}$^a$,
  Tobias Striegl\email{tobias.striegl}{physik.uni-siegen.de}$^b$}

\address{\it%
$^a$DESY Theory Group, \\
  Notkestr. 85, 22607 Hamburg, Germany \\[.5\baselineskip]
$^b$University of Siegen, \\
  Physics Department, Walter-Flex-Straße 3, 57068 Siegen, Germany \\[.5\baselineskip]
$^c$University of Würzburg, \\
  Faculty of Physics and Astronomy, Emil-Hilb-Weg 22, 97074 Würzburg, Germany
}

\abstract{
\centering
\begin{minipage}{0.845\textwidth}
This article summarizes the talk given at the LCWS 2021 conference on the status and news of the \whizard Monte Carlo event generator.
We presented its features relevant for the physics program of future lepton and especially linear colliders
as well as recent developments towards including NLO perturbative corrections and a UFO interface to study models beyond the Standard Model.
It takes as reference the version \texttt{3.0.0$\beta$} released in August 2020
and additionally discusses the developments that will be included in the next major version \texttt{3.0.0} to be released in April 2021.
\end{minipage}
\vskip 4.5cm
}
\maketitle

\section{About \whizard}
\whizard is a multi-purpose Monte Carlo event generator for hadron and lepton collider physics~\cite{Kilian:2007gr} providing all parts of event generation.
It is a general framework for all types of colliders, but with a special emphasis on the physics program of past and future lepton colliders.
\whizard has been applied in many lepton collider studies including the design reports for ILC, CLIC and FCC-ee~\cite{Fujii:2015jha,deBlas:2018mhx,Baer:2013cma,Behnke:2013lya,Abada:2019zxq}.\\
\whizard is shipped with a collection of subpackages.
For the generation of hard scattering matrix elements, \whizard allows to make use of its intrinsic tree-level matrix element generator \OMega~\cite{Moretti:2001zz} which applies the color-flow formalism for QCD~\cite{Kilian:2012pz}.\\
Besides \OMega, \whizard provides interfaces to the external matrix element generators \openloopstwo~\cite{Cascioli:2011va, Buccioni:2019sur}, \recola~\cite{Actis:2012qn, Actis:2016mpe} and \gosam~\cite{Cullen:2011ac, Cullen:2014yla} which can also be used to access loop matrix elements to include next-to-leading order perturbative corrections.\\
To integrate those matrix elements, \whizard originally employed the Monte Carlo integrator \vamp~\cite{Ohl:1998jn} which recently has been superseded by \vamptwo~\cite{Brass:2018xbv}, a complete re-implementation of the original \vamp algorithm parallelized via a message-passing interface (MPI).\\
Specifically for lepton collider physics, \whizard is able to take into account lepton collider beam spectra using the \circeone~\cite{Ohl:1996fi} and \circetwo subpackages.\\
Concerning the code base, \whizard is written in modern \texttt{Fortran 2008} and compilable by \texttt{gfortran} compilers newer than version \texttt{5.4.0}, \texttt{nagfor} compilers newer than version \texttt{6.2} and \texttt{ifort} compilers newer than version \texttt{18.0.0} while \OMega is written in \texttt{OCaml} supporting \texttt{OCaml} compiler versions newer than \texttt{4.02.3}.\\
\whizard's functionality is secured by a continuous integration system including roughly $500$ unit and functional tests hosted by the University of Siegen.

\section{Lepton collisions with \whizard}
\whizard is predominantly employed to study lepton collider physics for a number of reasons.
On the one hand, it offers the possibility to take into account non-trivial beam energy spectra.
Beam energy spectra at high energy lepton colliders are non-trivial for the reason that dense beam bunches are required to achieve high collider luminosities.
Dense beam bunches cause the strong electromagnetic fields of both bunches to influence each other leading to energy loss of the beams and thus of the colliding particles.
In \whizard, this is accounted for by the packages \circeone and \circetwo.
Both packages fit the energy spectrum obtained from \guineapig~\cite{Schulte:1997nga, Schulte:1999tx, Schulte:2007zz}.
While \circeone used a multi-parameter fit using a pre-defined function, \circetwo samples a two dimensional histogram in the energy fractions $x_1$ and $x_2$ from both beams that enter the hard interaction and then uses a Gaussian filter to remedy artifacts from limited \guineapig statistics.
Both versions of \circe thus generate collider specific files containing the fitted spectrum for use with \whizard.
For a large number of frequently studied collider setups, these files are available on the \href{https://whizard.hepforge.org/circe_files/}{\whizard HepForge page}\footnote{\url{https://whizard.hepforge.org/circe_files/}}. Alternatively, \whizard also allows to simulate simple Gaussian beam energy spreads.\\
On the other hand, the colliding partons lose energy due to bremsstrahlung also known as initial-state radiation (ISR).
Because of the small electron mass, this energy is predominantly converted into the emission of soft and collinear photons.
While soft photon emission can be resummed to all orders~\cite{Gribov:1971zn, Kuraev:1985hb}, the hard-collinear photons need to be resummed order by order.
In \whizard, this resummation is performed up to the leading-logarithmic order in the strict collinear limit of the emission~\cite{Skrzypek:1990qs}.
When generating events however, having purely collinear photon emission is not sufficient anymore.
To generate a non-trivial transverse-momentum spectrum for the emission, \whizard uses an \verb`isr_handler` which adds one photon with non-zero transverse momentum per beam to the event record and Lorentz-boosts the remaining event to reinstate energy-momentum conservation.
The number of photons as well as their tranverse-momentum distribution are currently heuristically motivated and could be improved in the future by taking into account higher orders of resummation.\\
The proposed future lepton colliders CLIC and ILC will offer the possibility to study collisions of polarized electrons with, in the case of ILC also polarized, positrons.
To keep up with these advancements in technology, \whizard supports event generation with arbitrary polarization setups including circular and longitudinal polarization and for massive colliding particles also transverse polarization with arbitrary polarization axes and fractions for both beams.
It is thus well suited for the simulation of lepton collisions at future machines.\\
In the past decades, the computational power of single cores rose almost linearly with time~\cite{rupp_karl_2020_3947824}.
This progression however slowed down in recent years while the computational effort spend on Monte Carlo event generators continued to increase.
To avoid studies beeing limited by the statistical uncertainties of Monte Carlo events without relying on increasing single-core performance, it is inevitable to provide parallelizable software able to run distributed over a large number of cores at the same time.
For this reason, \whizard's traditional doubly-adaptive multi-channel Monte Carlo integrator \vamp has been superseded by a fully MPI-parallelizable re-implementation of the original \vamp algorithm, dubbed \vamptwo.
This allows to achieve speedups of a factor of $10$ to $30$ and scales up to a usage of $\mathcal{O}\left( 100 \right)$ CPU cores making \whizard ready for applications on larger computing clusters.\\
Monte Carlo event generators are usually just one part in a larger toolchain.
The events generated by a fixed order Monte Carlo event generator still need to be processed by other tools for example detector simulations.
In order for this to be possible, events need to be exchanged between programs.
Besides direct interfaces, the most common option to interchange event records is to write them to disk.
\whizard supports exporting events in many different event formats.
In the basic version without any external tools, \whizard can output events in the formats \texttt{StdHEP}, \texttt{LHA}, \texttt{ascii}, \texttt{LHEF2} and \texttt{LHEF3}~\cite{Alwall:2006yp} as well as a number of different internal formats used for debugging purposes.
Additionally, \whizard can be linked to \texttt{LCIO}~\cite{Gaede:2003ip} or \texttt{HepMC2}~\cite{Dobbs:2001ck} or \texttt{HepMC3}~\cite{BUCKLEY2021107310} in order to export events in the corresponding formats including \texttt{ROOT}-trees~\cite{Brun:1997pa} via the \texttt{HepMC3} interface.
In summary, \whizard supports to simulate many lepton collider specific physics features and fits very well in the toolchains of large scale applications.

\section{\whizard \texttt{3.0}: Status and News}
\subsection{NLO QCD -- fixed order cross sections}
In the second quarter of 2021, the next major version \texttt{3.0.0} of \whizard will be released.
Its main innovation is the full support of perturbative NLO QCD corrections for total cross sections and fixed-order events at hadron and lepton colliders.\\
\whizard's journey to become an NLO Monte Carlo program started already more than a decade ago with hard-coded NLO electroweak corrections for chargino production~\cite{Kilian:2006cj, Robens:2008sa} at the ILC and hard-coded NLO QCD corrections to the production of two bottom-quark pairs at the LHC~\cite{Binoth:2009rv, Greiner:2011mp}.
A decade later, NLO QCD corrections have been implemented in a general and process independent way using the Frixione-Kunszt-Signer (FKS) subtraction scheme~\cite{Frixione:1995ms}.
All tree-level diagrams required for this computation can be provided by \OMega while virtual amplitudes are taken from an external one-loop provider such as \openloopstwo~\cite{Cascioli:2011va, Buccioni:2019sur} or \recola~\cite{Actis:2012qn, Actis:2016mpe}.
The construction of the subtraction terms, the book keeping of matrix elements required for the integration as well as the event generation are then performed internally in \whizard.\\
As a first step, we validated a large number of NLO QCD cross sections for processes common at lepton and hadron colliders by comparison with the Monte Carlo event generator \madgraph~\cite{Alwall:2014hca}.
For this, we chose a setup similar to the one chosen in~\cite{Frederix:2009yq}, namely
\begin{itemize} \setlength\itemsep{0mm}
  \item $\sqrt{s} = 1\TeV$ for lepton collider processes and $\sqrt{s} = 13\TeV$ for hadron collider processes,
  \item $m_h = 125\GeV$, $m_t = 173.2\GeV$, $\Gamma_Z = \Gamma_W = \Gamma_\text{top} = 0$,
  \item a diagonal CKM matrix,
  \item to compute all processes with explicit $b$-quarks in the final state in the four-flavor scheme using \texttt{MSTWnlo2008}~\cite{Martin:2009iq} PDFs with $n_f = 4$ and to use the five-flavor scheme and \texttt{MSTWnlo2008} PDFs with $n_f = 5$ otherwise,
  \item a scale given by $\mu = \mu_F = \mu_R = \nicefrac{H_T}{2} = \frac12 \sum_i \sqrt{p_{T,i}^2 + m_i^2}$ where the sum runs over all final-state particles,
  \item an anti-$k_T$ jet clustering algorithm~\cite{fastjetmanual, Cacciari:2008gp} with $R = 0.5$ and we required all Born jets to satisfy $p_T > 30\GeV$ and $\left\lvert\eta\right\rvert < 4$.
\end{itemize}
A selection of this endeavor's results are shown in Tab.~\ref{tab:validation_lepton} for lepton collider processes and Tab.~\ref{tab:validation_hadron} for hadron collider processes.
Among these processes, we computed cross sections for jet production at a lepton collider for up to six jets in the final state.
With each additional final-state jet, the amount of possible final-state-flavor combinations and the complexity of the virtual amplitudes grows tremendously.
While the cross section for \eejjjjj has been computed before in~\cite{Frederix:2009yq},
to our knowledge, this is the first time that the cross section for \eejjjjjj at a lepton collider has been computed with NLO QCD accuracy.
These fixed-order NLO QCD cross sections are publicly available from \whizard version \texttt{3.0.0}$\alpha$ on.

\begin{table}[H]
  \centering
  \begin{tabular}{l|lll|lll|l}
    \multicolumn{1}{c}{Process} & \multicolumn{3}{c}{\madgraph} & \multicolumn{3}{c}{\whizard} & \multicolumn{1}{c}{} \\
    & $\sigma_{\text{\textsc{lo}}}[\text{fb}]$ &
    $\sigma_{\text{\textsc{nlo}}}[\text{fb}]$ & $K$ & $\sigma_{\text{\textsc{lo}}}[\text{fb}]$ &
    $\sigma_{\text{\textsc{nlo}}}[\text{fb}]$ & $K$ & $\sigma_{\text{\textsc{nlo}}}^{\text{std}}$\\
    \hline \hline
    $e^+e^-\to jj$                & \tableline{622.70}{5}{639.30}{12}{}     & \tableline{622.737}{8}{639.39}{5}{}   &   \stddev{639.30}{0.12}{639.39}{0.05}\\
    $e^+e^-\to jjj$               & \tableline{340.4}{7}{317.3}{8}{}        & \tableline{340.6}{5}{317.8}{5}{}      &   \stddev{317.3}{0.8}{317.8}{0.5}\\
    $e^+e^-\to jjjj$              & \tableline{104.09}{20}{103.67}{26}{}    & \tableline{105.0}{3}{104.2}{4}{}      &   \stddev{103.67}{0.26}{104.2}{0.4}\\
    $e^+e^-\to jjjjj$             & \tableline{22.35}{5}{24.65}{4}{}        & \tableline{22.33}{5}{24.57}{7}{}      &   \stddev{24.65}{0.04}{24.57}{0.07}\\
    $e^+e^-\to jjjjjj$            & ~~~-- & ~~~-- & ~~~--                   & \tableline{3.583}{17}{4.46}{4}{}      &   ~~~-- \\
    \hline
    $e^+e^-\to t\bar{t}$          & \tableline{166.32}{11}{174.5}{3}{}      & \tableline{166.37}{12}{174.55}{20}{}  &   \stddev{174.5}{0.3}{174.55}{0.20}\\
    $e^+e^-\to t\bar{t}j$         & \tableline{47.95}{9}{53.336}{10}{}      & \tableline{48.12}{5}{53.41}{7}{}      &   \stddev{53.336}{0.010}{53.41}{0.07}\\
    $e^+e^-\to t\bar{t}jj$        & \tableline{8.608}{18}{10.515}{19}{}     & \tableline{8.592}{19}{10.526}{21}{}   &   \stddev{10.515}{0.019}{10.526}{0.021}\\
    $e^+e^-\to t\bar{t}jjj$       & \tableline{1.0371}{21}{1.415}{4}{}      & \tableline{1.035}{4}{1.405}{5}{}      &   \stddev{1.415}{0.004}{1.405}{0.005}\\
    \hline
  \end{tabular}
  \caption{Selection of validated lepton collider processes at LO and NLO QCD. Results from \madgraph were first computed in~\cite{Frederix:2009yq} but have been recomputed by us to include corrections and improvements made after the original paper has been published.}
  \label{tab:validation_lepton}
\end{table}
\begin{table}[H]
  \centering
  \begin{tabular}{l|lll|lll|l}
    \multicolumn{1}{c}{Process} & \multicolumn{3}{c}{\madgraph} & \multicolumn{3}{c}{\whizard} & \multicolumn{1}{c}{}\\
    & $\sigma_{\text{\textsc{lo}}}[\text{fb}]$ &
    $\sigma_{\text{\textsc{nlo}}}[\text{fb}]$ & $K$ & $\sigma_{\text{\textsc{lo}}}[\text{fb}]$ &
    $\sigma_{\text{\textsc{nlo}}}[\text{fb}]$ & $K$ & $\sigma_{\text{\textsc{nlo}}}^{\text{std}}$\\
    \hline \hline
    $pp \to jj$                   & \tableline{1.1593}{23}{1.6040}{29}{9} & \tableline{1.162}{4}{1.601}{5}{9}   & \stddev{1.6040}{0.029}{1.601}{0.005}\\
    $pp \to jjj$                  & \tableline{8.940}{21}{7.619}{19}{7} & \tableline{9.01}{4}{7.46}{9}{7}     & \stddev{7.619}{0.019}{7.46}{0.09}\\
    \hline
    $pp \to t\bar{t} $            & \tableline{4.584}{3}{6.746}{14}{5}  & \tableline{4.589}{9}{6.740}{10}{5}  & \stddev{6.746}{0.014}{6.740}{0.010}\\
    $pp \to t\bar{t} j$           & \tableline{3.133}{5}{4.095}{8}{5}   & \tableline{3.123}{6}{4.087}{9}{5}   & \stddev{4.095}{0.008}{4.087}{0.009}\\
    $pp \to t\bar{t} jj$          & \tableline{1.363}{3}{1.784}{3}{5}   & \tableline{1.360}{4}{1.775}{7}{5}   & \stddev{1.784}{0.003}{1.775}{0.007}\\
    $pp \to t\bar{t} t\bar{t}$    & \tableline{4.505}{5}{9.076}{13}{}   & \tableline{4.485}{6}{9.070}{9}{}    & \stddev{9.076}{0.013}{9.070}{0.009}\\
    \hline
    $pp \to t\bar{t} W^\pm$       & \tableline{3.777}{3}{5.668}{18}{2}  & \tableline{3.775}{5}{5.674}{5}{2}   & \stddev{5.668}{0.018}{5.674}{0.005}\\
    $pp \to t\bar{t} W^\pm j$     & \tableline{2.352}{3}{3.434}{8}{2}   & \tableline{2.356}{7}{3.427}{8}{2}   & \stddev{3.434}{0.008}{3.427}{0.008}\\
    $pp \to t\bar{t} Zj$          & \tableline{3.953}{4}{5.079}{14}{2}  & \tableline{3.943}{14}{5.069}{17}{2} & \stddev{5.079}{0.014}{5.069}{0.017}\\
    $pp \to t\bar{t} ZZ$          & \tableline{1.349}{14}{1.843}{4}{}   & \tableline{1.3590}{29}{1.842}{3}{}  & \stddev{1.843}{0.004}{1.842}{0.003}\\
    $pp \to HZ$                   & \tableline{6.468}{8}{7.693}{19}{2}  & \tableline{6.474}{11}{7.679}{12}{2} & \stddev{7.693}{0.019}{7.679}{0.012}\\
    \hline
  \end{tabular}
  \caption{Selection of validated hadron collider processes at LO and NLO QCD. Results from \madgraph were first computed in~\cite{Frederix:2009yq} but have been recomputed by us to include corrections and improvements made after the original paper has been published.}
  \label{tab:validation_hadron}
\end{table}

\subsection{NLO QCD -- fixed order differential distributions}
\label{sec:fNLOdistributions}
\begin{figure}[H]
  \centering
  \subfloat{
  \includegraphics[width=0.5\textwidth]{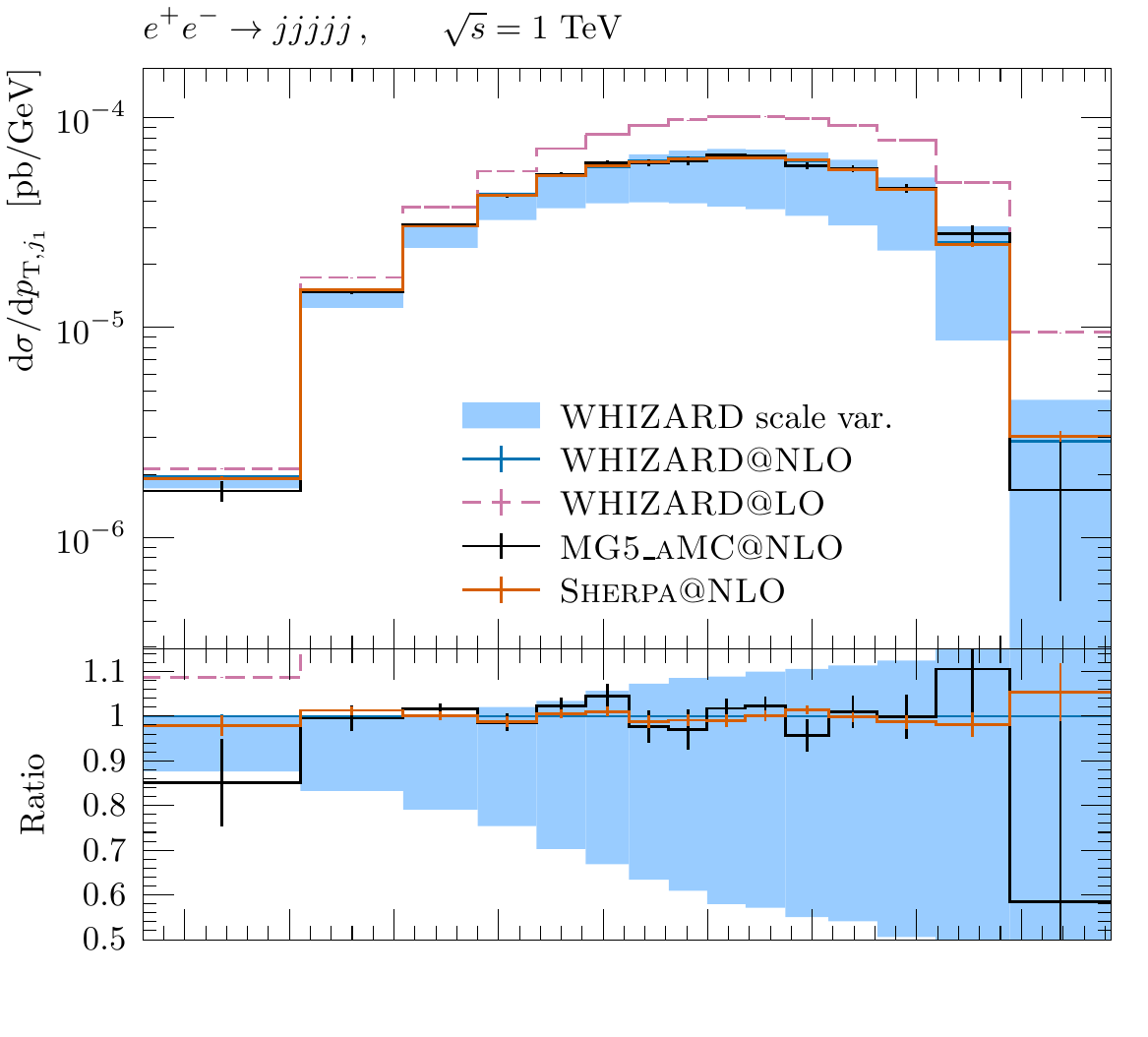}}
  \subfloat{
  \includegraphics[width=0.5\textwidth]{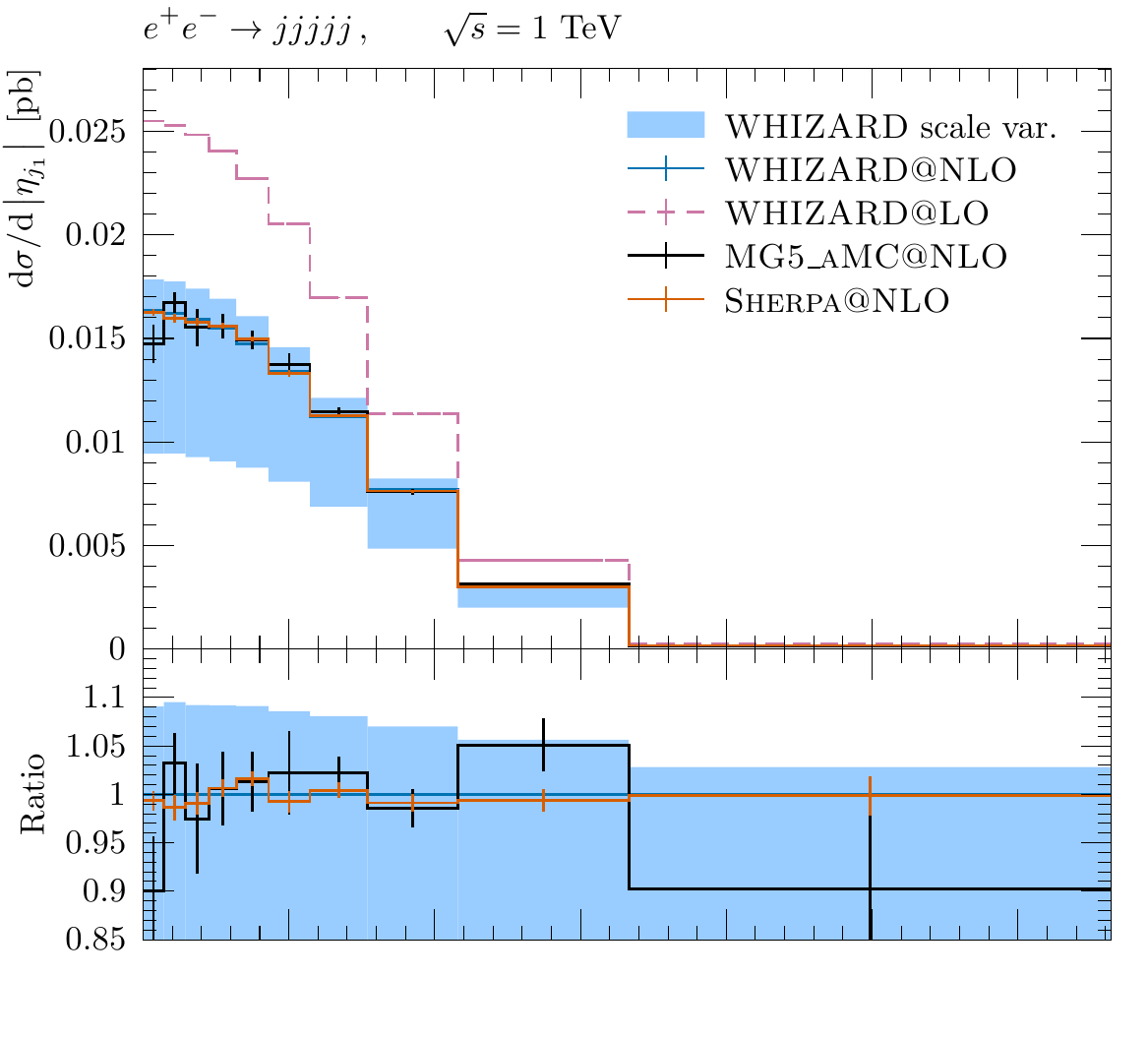}}\\[-44.6pt]
  \setcounter{subfigure}{0}
  \subfloat[Transverse momentum distribution of the hardest jet]{\label{fig:4jPt1}
  \includegraphics[width=0.5\textwidth]{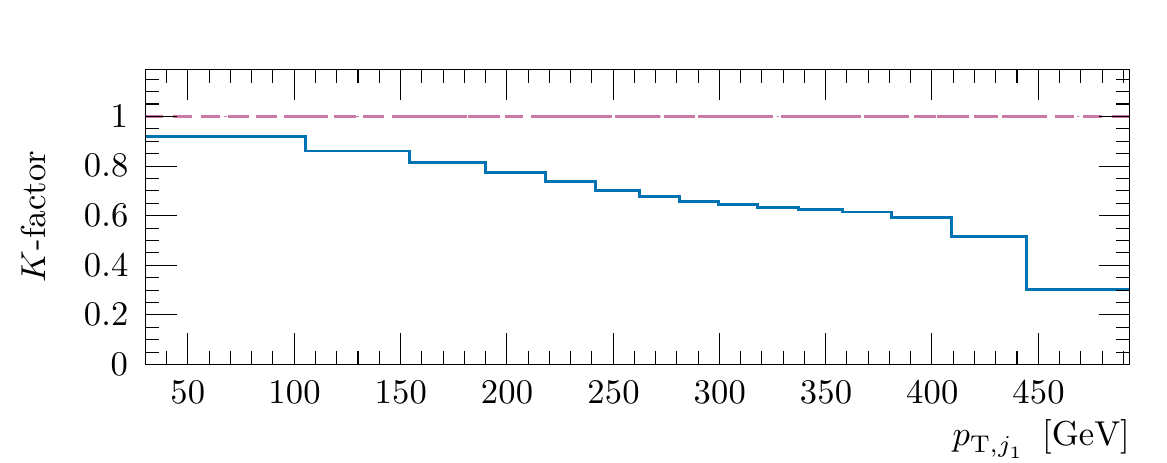}}
  \subfloat[Rapidity distribution of the hardest jet]{\label{fig:4jEta2}
  \includegraphics[width=0.5\textwidth]{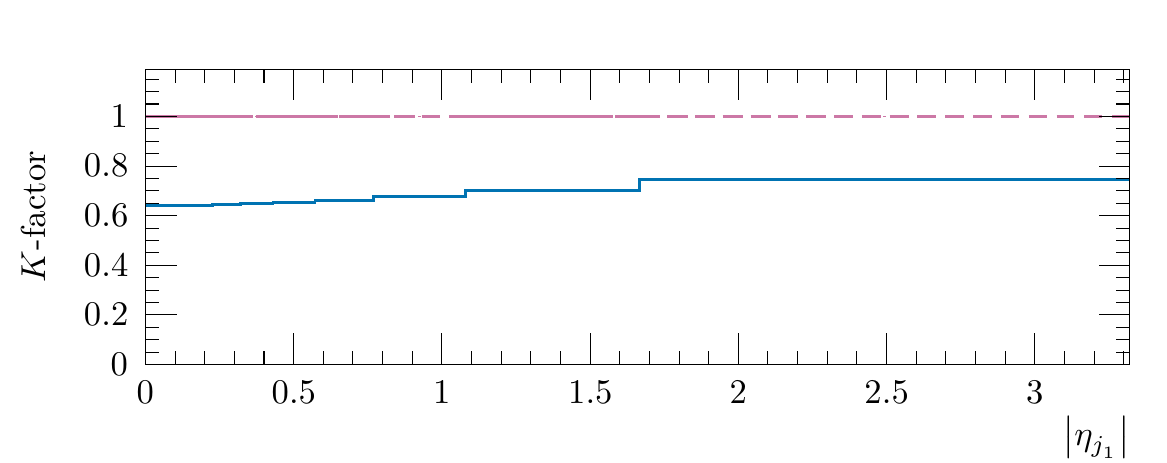}}\\
  \caption{Comparison of kinematic distributions of the hardest jet in \eejjjjj at fixed-order NLO QCD between \madgraph, \sherpa and \whizard. The vertically centered subplot shows the ratio over \whizard's NLO distribution while the lower subplot shows the $K$ factor given by $\nicefrac{\sigma_\text{NLO}}{\sigma_\text{LO}}$. The scale uncertainty bands are obtained by varying the scale $\mu$ by a factor of $2$ up and down. For \sherpa and \whizard, $5\cdot10^8$ event groups have been generated while for \madgraph, a factor of $5$ less event groups contribute.}
  \label{fig:5jPt1Eta1}
\end{figure}

In the most recently released version \texttt{3.0.0}$\beta$ of \whizard, the NLO QCD capabilities have been extended from fixed-order cross sections to fixed-order differential distributions.
So, from version \texttt{3.0.0}$\beta$ onwards, \whizard is able to generate differential distributions at fixed NLO QCD for both, leptonic and hadronic initial states.
For example, Fig.~\ref{fig:5jPt1Eta1} shows the distributions of the transverse momentum and the absolute value of the rapidity for the hardest jet in \eejjjjj at LO and NLO from \whizard compared with the NLO results from \madgraph and \sherpa.
For this comparison at the level of events, a similar setup to the one used for the fixed-order cross sections has been used.
The only differences to the setup described before are
\begin{itemize} \setlength\itemsep{0mm}
  \item a fixed scale choice of $\mu = \mu_F = \mu_R = m_Z$ to avoid effects of minor differences in the running of the strong coupling in \madgraph, \sherpa and \whizard
  \item the usage of the generalized $k_t$ jet clustering algorithm~\cite{fastjetmanual, CATANI1991432, PhysRevD.48.3160} with $R = 0.5$ and $p = -1$ which is better suited for $e^+ e^-$ collisions.
\end{itemize}
As a conclusion, we observe very good agreement between the different Monte Carlo event generators also at the level of differential distributions.

\subsection{NLO QCD -- top-pair production at the threshold}
\begin{figure}[H]
  \centering
  \begin{minipage}{0.49\textwidth}
    \centering
    \subfloat[Matched total production cross section]{
      \includegraphics[width=0.95\textwidth]{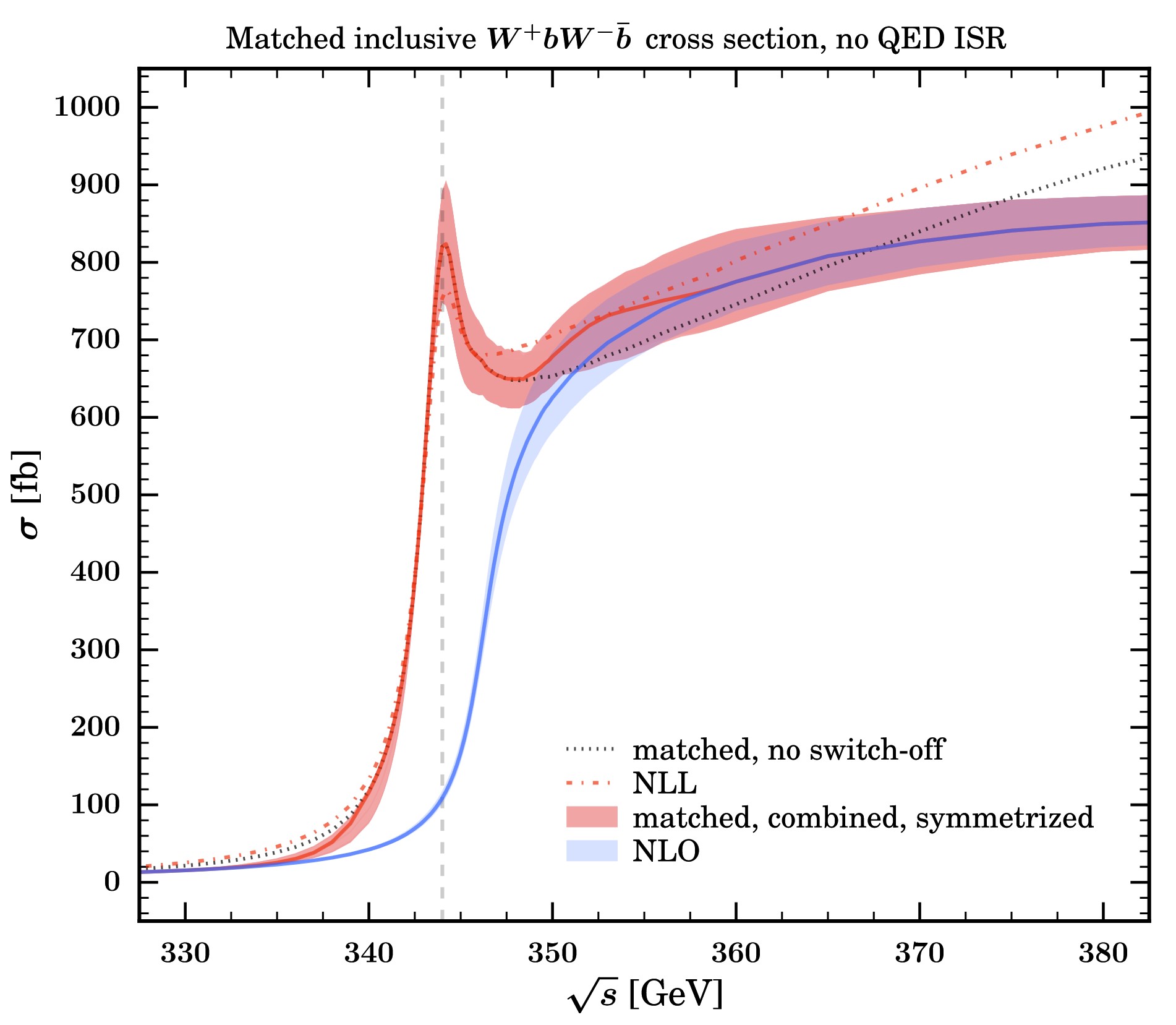}
      \label{fig:thresholdscan}
    }
  \end{minipage}\hfill
  \begin{minipage}{0.49\textwidth}
    \centering
    \subfloat[Invariant mass distribution of the reconstructed top quark at $\sqrt{s} = 344\GeV$]{
      \includegraphics[width=0.95\textwidth]{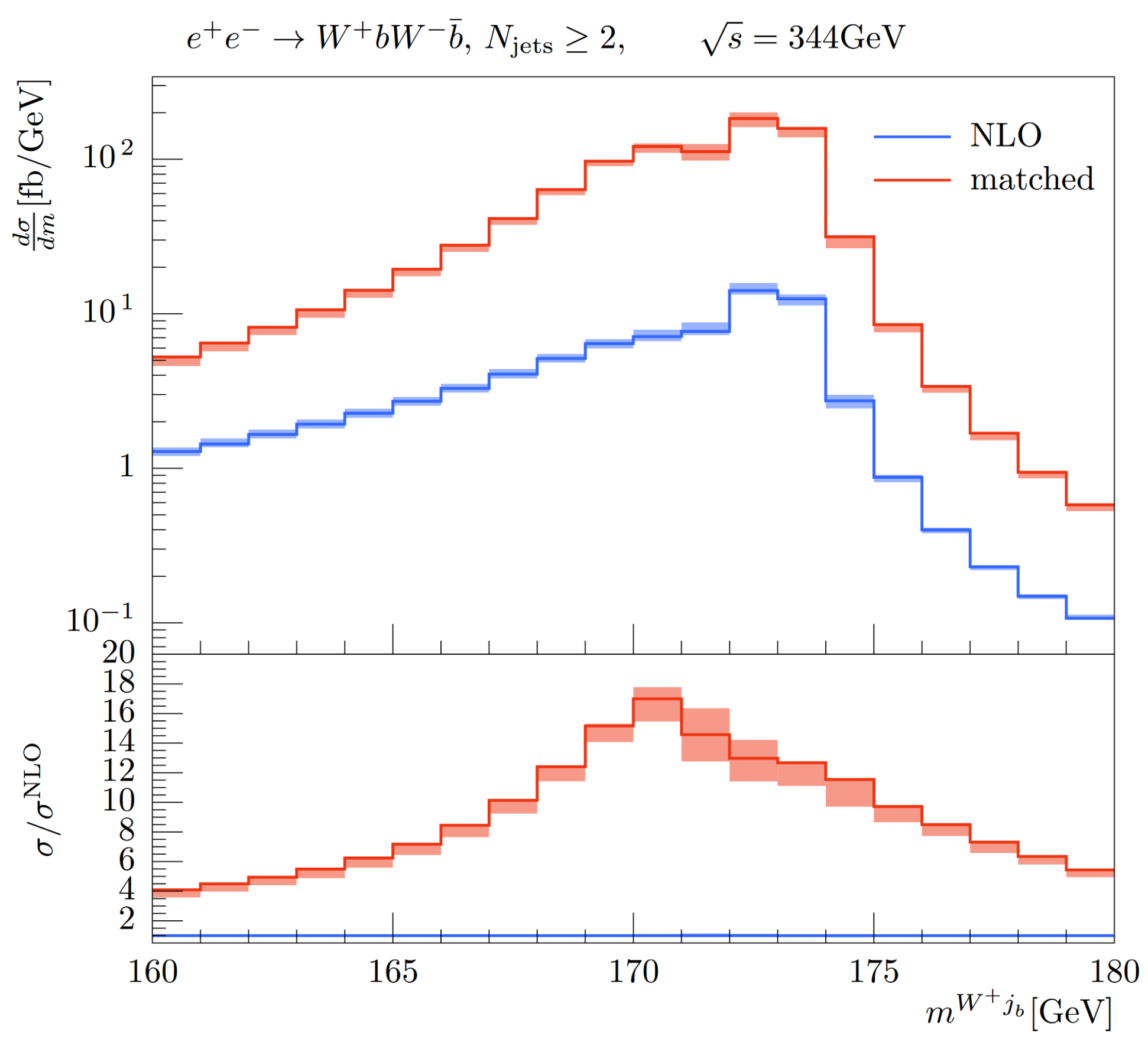}
      \label{fig:thresholdmwj}
    }
  \end{minipage}
  \captionof{figure}{Total production cross section and the distribution of the invariant mass of the reconstructed top quark consisting of the $W^+$ and the $b$-jet in $e^+e^- \to W^+ b W^- \bbar$.  The blue line shows the fixed-order NLO QCD predicsion while the red line shows the matched predictions at NLO QCD+NLL. The bands depict the full combination of symmetrized scale variations and for the matched curve also the matching parameter variations. Taken from~\cite{Bach:2017ggt}.}
  \label{fig:threshold}
\end{figure}

One more example application of \whizard's NLO QCD corrections is the inclusion of non-relativistic effects to the top-anti-top threshold resummation at NLL matched to NLO QCD corrections as described in~\cite{Bach:2017ggt}.
A selection of the main results is shown in Fig.~\ref{fig:threshold}.
It depicts in Fig.~\ref{fig:thresholdscan} the total production cross sections of exclusive $W^+ b W^- \bbar$ production at a lepton collider for varying center-of-mass energies $\sqrt{s}$ for both, the fixed-order prediction as shown in blue and the predictions including $t\tbar$ form factors for the NLL threshold resummation derived in non-relativistic QCD matched to the full fixed-order QCD prediction valid in the continuum.
All of these predictions are implemented and available in \whizard and thus allow to study also differential observables with the same accuracy.
For example, Fig.~\ref{fig:thresholdmwj} shows the distribution of the invariant mass of the reconstructed top quark around the top pole.
Taking these effects into account is crucial for the top-mass determination at any future lepton collider running at a centre-of-mass energy close to the top threshold and for the assessment of experimental systematic uncertainties.

\subsection{\whizard beyond the Standard Model}
Besides the work on including NLO QCD corrections in our predictions, we also aim to allow making LO predictions for arbitrary physics models beyond the Standard Model with \whizard.
\whizardtwo already offers a wide range of efficiently hard-coded models ranging from different simplified or more detailed versions of the Standard Model including non-diagonal CKM matrix elements or Higgs couplings for light fermions over simple extensions of the Standard Model such as the inclusion of a $Z^\prime$ Boson or the Two-Higgs-Doublet model to various flavors of the Little(st) Higgs model.
From version \texttt{2.8.5} on, \whizard additionally offers the possibility to compute cross sections and generate events at the LO for any model formulated in the Universal Feynman Output (UFO) format~\cite{Degrande:2011ua} as provided e.g. by \textsc{FeynRules}~\cite{Alloul:2013bka} and \texttt{SARAH}~\cite{Staub:2015kfa}.
The matrix element code generated from these models is not as highly optimized but these models are more flexible than the hard-coded ones.
So far, this includes models with particles with arbitrary spin values, arbitrary Lorentz structures, up to 5- and 6-point vertices and also BSM models given in the SLHA input format~\cite{Skands:2003cj, Allanach:2008qq}.
Models with Majorana fermions however will only fully be supported from version \texttt{3.0.0} onwards also fully supporting the UFO extensions from \texttt{SMEFTsim} 3.0~\cite{Brivio:2020onw}.
Additionally, version \texttt{3.0.0} will include a number of technical bug fixes to tolerate SLHA 2 QNUMBER blocks and particles with negative particle ID and to allow more flexible propagator definitions as well as fractional hypercharges.

\subsection{\whizard beyond version \texttt{3.0.0} -- \powheg matching}
The differential distributions shown in Sec.~\ref{sec:fNLOdistributions} are only accurate up to fixed order in NLO QCD.
In the case of \eejjjjj, this means that all the events that pass through the analysis either feature $5$ or $6$ QCD particles in the final state which then are clustered into up to $6$ jets.
Although this prescription is theoretically well defined, in order to achieve agreement with experimental measurements, a resummation of large logarithms stemming from soft and collinear QCD emissions in form of a parton shower is missing.
However, if we applied a parton shower naively to the fixed-order events, we would take into account diagrams with one more final-state QCD parton than present at LO described by both, the hard matrix element as well as the parton shower, so twice in total.
We need to apply a proper matching scheme to circumvent this double counting of contributions.
There are several different schemes known in the literature for matching an NLO prediction to a parton shower, most notably \texttt{MC@NLO}~\cite{Alwall:2014hca} and \powheg matching~\cite{Nason:2004rx, Frixione:2007vw}.
In \whizard, we decided to implement the \powheg matching as it fits best to how NLO matrix elements are organized in the FKS subtraction scheme.
The name \powheg stands for \emph{po}sitive-\emph{w}eight \emph{h}ardest \emph{e}mission \emph{g}enerator and as this suggests, \powheg matching also has the advantage of yielding matched NLO events without any negative weights which is very desirable for processing the event weights in statistical analyses afterwards.
The work to implement \powheg matching in the Monte Carlo generator \whizard has already started a few years ago~\cite{Nejad:2015opa} but has so far been validated only for the processes \eett and \eetth.
The generalization of the \powheg matching to arbitrary processes is one of the projects we are working on which will be released in one of the versions after \whizard \texttt{3.0.0}.

\subsection{\whizard beyond version \texttt{3.0.0} -- NLO EW}
The second project we are currently working on is the inclusion of next-to-leading order electroweak corrections in predictions for cross sections as well as differetial distributions.
There are several things to do for the transition from NLO QCD corrections to NLO EW corrections.
First of all, while for NLO QCD we allowed gluon emissions off quarks and gluon splittings into quarks to construct the possible real flavor structures based on a given Born flavor structure, we replaced these by photon emissions off charged fermions and photon splittings for real NLO QED corrections.
We also implemented the corresponding electroweak couplings and splitting functions for the construction of the electroweak subtraction terms.
For the full NLO EW corrections, we additionally had to take into account $W$ and $Z$ and Higgs bosons in the virtual amplitudes.\\
There are still a number of things that need to be taken care of.
These include the correct treatment of the photon content in the proton PDFs and the implementation of algorithms for electron-photon recombination to guarantee infrared-safe observables.
As a next step, we aim to allow for electron PDFs to access the photon content of the electron~\cite{Frixione:2019lga, Bertone:2019hks} and an infrastructure for mixed coupling expansions to enable predictions with combined NLO QCD and NLO EW accuracy.\\
Nevertheless, a number of processes shown in Tab.~\ref{tab:nloewcrosssections} could already be validated.
For $pp \to ZZZ$, we chose a setup similar to the one chosen in~\cite{Frederix:2018nkq} while the setup for the remaining two processes differs slightly in some parameter choices.
The settings that differ from~\cite{Frederix:2018nkq} are the choice of
\begin{itemize} \setlength\itemsep{0mm}
  \item a fixed scale $\mu = m_Z$,
  \item heavy boson masses of $m_Z = 91.1882 \GeV$ and $m_W = 80.385 \GeV$
  \item and particle widths of $\Gamma_Z = 2.4955 \GeV$ and $\Gamma_W = 2.0897 \GeV$
\end{itemize}
for both, \madgraph and \whizard.

\begin{table}[H]
  \centering
  \begin{tabular}{l|ll|ll|ll}
    \multicolumn{1}{c}{Process} & \multicolumn{2}{c}{\madgraph} & \multicolumn{2}{c}{\whizard} & \multicolumn{2}{c}{}\\
    & $\sigma_{\text{\textsc{lo}}}[\text{fb}]$ & $\sigma_{\text{\textsc{nlo}}}[\text{fb}]$
    & $\sigma_{\text{\textsc{lo}}}[\text{fb}]$ & $\sigma_{\text{\textsc{nlo}}}[\text{fb}]$
    & $\nicefrac{\sigma_{\text{\textsc{nlo}}}}{\sigma_{\text{\textsc{lo}}}}$ & $\sigma_{\text{\textsc{nlo}}}^{\text{std}}$\\
    \hline \hline
    $pp \to ZZZ$                & $1.0761(1)  \cdot 10^1$ & $0.9741(1) \cdot 10^1$     & $1.0752(8) \cdot 10^1$ & $0.9727(9) \cdot 10^1$     & $0.905$ & $1.00$ \\
    $pp \to \nu_e \nubar_e$     & $3.2947(4)  \cdot 10^6$ & $3.3136(4) \cdot 10^6$     & $3.2949(10) \cdot 10^6$ & $3.3138(10) \cdot 10^6$   & $1.006$ & $0.02$ \\
    $pp \to e^+ \nu_e$          & $1.78224(13)  \cdot 10^7$ & $1.77598(15) \cdot 10^7$ & $1.78224(21) \cdot 10^7$ & $1.77621(21) \cdot 10^7$ & $0.997$ & $0.93$ \\
    \hline
  \end{tabular}
  \caption{Selection of validated hadron collider cross sections at NLO EW. \madgraph results partially taken from~\cite{Frederix:2018nkq}.}
  \label{tab:nloewcrosssections}
\end{table}

\section{Summary}
These proceedings document the advantages of using \whizard as a Monte Carlo event generator for computing cross sections and simulating events at future lepton colliders as well as the new features that will be contained in the soon-to-be released version \texttt{3.0.0} as depicted in a talk given at the LCWS 2021 conference.
These features mainly include the full support of NLO QCD cross sections and event generation at fixed order as well as the full support of the UFO format to specify models beyond the Standard Model.
It also contains a glimpse on the developments beyond version \texttt{3.0.0} which are the support for \powheg matching to match NLO QCD events to a parton shower for general processes as well as the inclusion of NLO EW corrections in the predictions of cross sections and events.

\section*{Acknowledgments}
This work was funded by the Deutsche Forschungsgemeinschaft under Germany’s Excellence Strategy – EXC 2121 "Quantum Universe" - 390833306.
PS also wants to thank the organizers of the LCWS 2021 for this interesting workshop covering a lot of different topics and an unusually large range of time and space each day.

\printbibliography

\end{document}